\documentclass[iop, aas_macros,numberedappendix,appendixfloats]{emulateapj}
\usepackage{graphicx}
\usepackage{epstopdf}
\usepackage[usenames,dvipsnames]{color}
\usepackage{amsmath,times}
\usepackage{float}
\usepackage{ulem}

\usepackage{subfigure}
\usepackage[export]{adjustbox}

\usepackage[bookmarks=false]{hyperref}
\hypersetup{
  colorlinks= true, 
  urlcolor  = Blue, 
  filecolor=black,
  runcolor=blue,
  linkcolor = red, 
  citecolor = blue 
} 
\newcommand{\ci}{C\,{\sc i}}

\newcommand{\lens}{2MASX J20295548+0120580}

\shorttitle{Multiwavelength characterisation of a DSFG}
\shortauthors{G. W. Roberts-Borsani et al.}

\begin{document}

\title{Multiwavelength characterisation of an ACT-selected, lensed dusty star-forming galaxy at $z$\,=\,2.64}

\author{G.~W. Roberts-Borsani,$^{1}$ M.~J. Jim\'enez-Donaire,$^{2}$ M. Dapr\`a,$^{3}$ K. Alatalo,$^{4}$\altaffilmark{$\dagger$} I. Aretxaga,$^{5}$ J. \'Alvarez-M\'arquez,$^{6}$ \\A.~J. Baker,$^{7}$ S. Fujimoto,$^{8}$ P.~A. Gallardo,$^{9}$ M. Gralla,$^{10}$ M. Hilton,$^{11}$ J.~P. Hughes,$^{7}$ C. Jim\'enez,$^{12}$ N. Laporte,$^{1}$ \\T.~A. Marriage,$^{13}$ F. Nati,$^{14}$ J. Rivera,$^{7}$ A. Sievers,$^{15}$ A. Wei$\ss$,$^{16}$ G.~W. Wilson,$^{17}$ E.~J. Wollack,$^{18}$ M.~S. Yun$^{17}$}

\affil{$^{1}$Department of Physics and Astronomy, University College London, Gower Street, London WC1E 6BT, UK\\
$^{2}$Institut f\"ur theoretische Astrophysik, Zentrum f\"ur Astronomie der Universit\"at Heidelberg, Albert-Ueberle Str. 2, D-69120 Heidelberg, Germany\\
$^{3}$Department of Physics and Astronomy, LaserLaB, VU University, De Boelelaan 1081, 1081 HV Amsterdam, the Netherlands\\
$^{4}$The Observatories of the Carnegie Institution for Science, 813 Santa Barbara St., Pasadena, CA 91101, USA\\
$^{5}$Instituto Nacional de Astrof\'isica \'Optica y Electr\'onica, Calle Luis Enrique Erro 1, Sta. Mar\'ia Tonantzintla, Puebla, M\'exico\\
$^{6}$Centro de Astrobiolog\'ia (CAB, CSIC-INTA), 28850 Torrej\'on de Ardoz, Madrid, Spain\\
$^{7}$Department of Physics and Astronomy, Rutgers, The State University of New Jersey, 136 Frelinghuysen Road, Piscataway, NJ 08854-8019, USA\\
$^{8}$Institute for Cosmic Ray Research, The University of Tokyo, Kashiwa-no-ha, Kashiwa 277-8582, Japan\\
$^{9}$Department of Physics, Cornell University, Ithaca, NY 14853, USA\\
$^{10}$Steward Observatory, University of Arizona, 933 North Cherry Avenue, Tucson, AZ 85721, USA\\
$^{11}$Astrophysics \& Cosmology Research Unit, School of Mathematics, Statistics, \& Computer Science, University of KwaZulu-Natal, Westville Campus, Durban 4041, ZA\\
$^{12}$Departamento de Astrof\'isica, Universidad de La Laguna, E-38206 La Laguna, Tenerife, Spain\\
$^{13}$Department of Physics and Astronomy, The Johns Hopkins University, 3400 N. Charles St., Baltimore, MD 21218-2686, USA\\
$^{14}$Department of Physics and Astronomy, University of Pennsylvania, 209 South 33rd Street, Philadelphia, PA, USA 19104\\
$^{15}$IRAM, Pico Veleta, Granada, Spain\\
$^{16}$Max-Planck-Institut f\"{u}r Radioastronomie, Auf dem H\"ugel 69 D-53121 Bonn, Germany\\
$^{17}$Department of Astronomy, University of Massachusetts, Amherst, MA 01003, USA\\
$^{18}$NASA/Goddard Space Flight Center, Greenbelt, MD 20771, USA
}

\altaffiltext{$\dagger$}{Hubble fellow}
\email{guidorb@star.ucl.ac.uk}

\begin{abstract}
We present \ci\,(2--1) and multi-transition $^{12}$CO observations of a dusty star-forming galaxy, ACT\,J2029+0120, which we spectroscopically confirm to lie at $z$\,=\,2.64. We detect CO(3--2), CO(5--4), CO(7--6), CO(8--7), and \ci\,(2--1) at high significance, tentatively detect HCO$^{+}$(4--3), and place strong upper limits on the integrated strength of dense gas tracers (HCN(4--3) and CS(7--6)). Multi-transition CO observations and dense gas tracers can provide valuable constraints on the molecular gas content and excitation conditions in high-redshift galaxies. We therefore use this unique data set to construct a CO spectral line energy distribution (SLED) of the source, which is most consistent with that of a ULIRG/Seyfert or QSO host object in the taxonomy of the \textit{Herschel} Comprehensive ULIRG Emission Survey. We employ RADEX models to fit the peak of the CO SLED, inferring a temperature of T$\sim$117 K and $n_{\text{H}_2}\sim10^5$ cm$^{-3}$, most consistent with a ULIRG/QSO object and the presence of high density tracers. We also find that the velocity width of the \ci\ line is potentially larger than seen in all CO transitions for this object, and that the $L'_{\rm C\,I(2-1)}/L'_{\rm CO(3-2)}$ ratio is also larger than seen in other lensed and unlensed submillimeter galaxies and QSO hosts; if confirmed, this anomaly could be an effect of differential lensing of a shocked molecular outflow.
\end{abstract}

\clearpage

\section{Introduction}
\label{sec:intro}
A longstanding problem in galaxy evolution is tracking the build-up of stellar mass over cosmic time (\citealt{madau2014}, and references therein). Gaining a complete census of the star formation history of the universe relies on a large suite of complementary observations, which can probe different regimes in star formation, including both unobscured (traced by ultraviolet light and nebular narrow emission lines, such as H$\alpha$) and obscured (traced by dust continuum emission) star formation. The discovery and follow-up of 850 $\mu$m-bright submillimeter galaxies (SMGs; \citealt{smail1997, barger98, hughes98}) with flux densities from a few to $\sim$10 mJy by the Submillimetre Common User Bolometric Array (SCUBA) opened a new window in mapping the buildup of stellar mass in the universe. The past decade has seen significant progress in deep, blank-field surveys at millimeter and submillimeter wavelengths over small patches of the sky. While rare, SMGs account for a non-negligible fraction of the cosmic star formation rate at $z$\,\textgreater\,1 \citep{blain2002,magnelli11,murphy11,casey12a,casey12b,magnelli13,casey2014}.

Until recently, the properties of SMGs were difficult to constrain, due to the lack of ultraviolet or optical counterparts with which their redshifts could be measured.  These starbursting galaxies are hypothesized to be the high-redshift progenitors of massive elliptical galaxies, but display little to no emission at shorter wavelengths. Early work aiming to localize SMGs using radio counterparts \citep{chapman2005} had limited success, for a possibly biased sample. More recently, the onset of high sensitivity, large bandwidth millimeter and submillimeter receivers has revolutionized the field of detecting and studying a variety of dusty star-forming galaxy (DSFG) populations (\citealt{casey2014}, and references therein), allowing for the determination of the redshifts for a substantial number of objects using bright molecular and atomic fine structure lines (e.g.,\citealt{weiss2009,vieira2013,weiss13,strandet16}). Despite this new discovery space opening for the study of SMGs, many questions remain about the state of their large reservoirs of molecular gas, including the characterisation of their kinematics \citep{hodge2012}, excitations \citep{walter2011}, and star formation efficiencies.

In this paper, we present new observations of a lensed DSFG detected 
by the Atacama Cosmology Telescope (ACT; \citealt{swetz2011}), 
ACT\,J2029+0120 (hereafter ACT\,J2029), using the Institut de 
Radioastronomie Millim\'etrique (IRAM) 30\,m telescope on Pico 
de Veleta. We use these new data to establish the galaxy's redshift, 
characterize the excitation conditions and other properties 
of its molecular gas, and confirm its status as a lensed 
system. In Section \S\ref{sec:obs} we describe our earlier observations of the 
source, which were insufficient 
to determine its redshift, and the new observations 
taken at the IRAM 30\,m. In Section \S\ref{sec:results} we present the parameters derived 
from these new observations, and in Section \S\ref{sec:disc} we discuss the implications 
of our observations for ACT\,J2029's intrinsic properties. The 
cosmological parameters $H_0 = 70\,{\rm km\,s^{-1}\,Mpc^{-1}}$, 
$\Omega_m = 0.3$, and $\Omega_\Lambda = 0.7$ are assumed throughout. All magnitudes are in the AB system \citep{okegunn83}.

\section{Observations}
\label{sec:obs}
\subsection{Atacama Cosmology Telescope}
\label{sec:act}
ACT\,J2029 was initially detected as a point source in ACT three-band 
observations of an equatorial field overlapping the Stripe 82 
region of deep Sloan Digital Sky Survey (SDSS) coverage (Gralla et al., in prep). 
Without deboosting corrections, its measured flux densities at 
148, 218, and 277\,GHz are $9.0 \pm 2.0\,{\rm mJy}$, $22.0 \pm 
3.0\,{\rm mJy}$, and $52.8 \pm 6.1\,{\rm mJy}$, respectively, 
with a nominal 218\,GHz position of $\alpha$=20:29:55.7 $\delta$=+01:20:54.5 (J2000; $\pm 
6^{\prime\prime}$ positional uncertainty). 

\subsection{Large Millimeter Telescope}
The source was further observed on 2013 May 30 with the Redshift Search Receiver (RSR; \citealt{erickson07}) on the Large Millimeter Telescope (LMT), with the aim of constraining its redshift (PI: Wilson). The total on-source integration time was 27,000\,s, with a system temperature $\sim 83\,{\rm K}$ and an elevation $\sim 60^\circ$ at which the LMT's elevation gain changes slowly. The data were calibrated following the standard procedures described by \citet{zavala15}. A single emission line was detected at a frequency of $95.0031 \pm 0.0015\,{\rm GHz}$, with an apparent flux of $6.05 \pm 0.28\,{\rm Jy\,km\,s^{-1}}$ and a velocity FWHM (corrected for instrumental resolution) of $236.6 \pm 9.1\,{\rm km\,s^{-1}}$. Given a pointing offset of $6.3^{\prime\prime}$ from the location of the peak CO emission (see Section \S\ref{sec:carma_sec} below) and together with the $23.4^{\prime\prime}$ effective width of the RSR beam at 95\,GHz (interpolating between 75 and 110\,GHz: \citealt{zavala15}), we corrected the apparent flux to $7.40 \pm 0.34 (\pm 0.74)\,{\rm Jy\,km\,s^{-1}}$ (including a 10\% uncertainty in the flux scale).
The absence of any other lines in the RSR's 73--111\,GHz bandwidth ruled out the CO(5--4) transition as the detected emission line, or any higher-$J$ transition. However, CO(4--3) at $z = 3.85$, 
CO(3--2) at $z = 2.64$, CO(2--1) at $z = 1.43$, and CO(1--0) at $z = 0.21$ remained as possible identifications.

\subsection{CARMA}
\label{sec:carma_sec}
ACT\,J2029 was re-observed with the Combined Array for Research in Millimeter-wave Astronomy (CARMA) in its compact D configuration at 95\,GHz (proposal c1204; PI Baker). The observations were taken on 2014 January 20 with a total of 4.8\,hrs on source. The data were calibrated with observations of J2134$-$018 (gain), J2015+372 (bandpass), and MWC349 (flux, with a reference flux density of 1.18\,Jy) following the procedures described in Appendix A of \citet{su15}, for a synthesized beam of $5.6^{\prime\prime} 
\times 4.4^{\prime\prime}$ at a position angle $80.8^\circ$. 
An integrated line spectrum was derived by taking the integrated intensity (moment0) map and summing over all pixels known to contain flux, as described in Section 3 of \citet{alatalo13}. The bandwidth over which we measured the line flux is 250\,MHz ($\approx$\,780\,km~s$^{-1}$). Upon creating the cube, we measure a 1$\sigma$ root mean square noise of 4.48 mJy~beam$^{-1}$ in 20\,km~s$^{-1}$ channels. Integrating over all channels with line emission (spanning 
$400\,{\rm km\,s^{-1}}$ in velocity), we obtained a spatially 
unresolved detection of line emission at an RA and DEC of $\alpha$=20:29:55.495 $\delta$=+01:20:58.944 (J2000). This is shown in Figure \ref{fig:lens}. We also attempted to detect 95\,GHz continuum by integrating over 7.25\,GHz 
of line-free channels, resulting in a 3$\sigma$ upper limit of 0.64\,mJy~beam$^{-1}$.

\subsection{Southern African Large Telescope}
The CARMA position of ACT\,J2029 revealed an apparent optical 
counterpart for the DSFG at $\alpha$=20:29:55.479 $\delta$=+01:20:58.610, \lens, for which optical spectroscopy at the 
Southern African Large Telescope \citep[SALT;][]{buckley+06} was obtained. Longslit observations were carried out using the Robert Stobie Spectrograph \citep[RSS,][]{burgh+03,kobulnicky+03} under program
2014-1-RU{\_}RSA-002 (PI: Hughes) on 2014 June 24. The PG0900 grating was used with two different grating angles (14$^\circ$ and 14.375$^\circ$) in order to cover the gaps between the camera's CCD chips. The total exposure time was 1200\,s, divided into two 300\,s exposures at each grating setting to aid in removing cosmic rays. The wavelength coverage was $\sim$3800 \AA-6900 \AA\ and the delivered spectral resolution was $\sim$950 at the central wavelength given the 1.5$^{\prime\prime}$ wide slit used. Stellar images on the acquisition frame were $\sim$2.2$^{\prime\prime}$ (FHWM) in size.

The data were reduced using the SALT science pipeline \citep{crawford+10}, and subsequent processing was carried out
using IRAF\footnote{
  IRAF is distributed by the National Optical
  Astronomy Observatory, which is operated by the Association of
  Universities for Research in Astronomy (AURA) under cooperative
  agreement with the National Science Foundation.}
tasks.
Cosmic rays were removed using L.A.Cosmic \citep{vandokkum01}. The
2D Argon arc-lamp images were used to determine the wavelength
calibration as well as the transformation necessary to rectify (i.e.,
straighten out) the wavelength solution along the imaging
direction. The accuracy of the wavelength solution on the arc images
was better than 0.5 \AA.  The wavelength solutions of the extracted
galaxy spectra were verified to be better than $\pm$20 km s$^{-1}$ by
checking the wavelengths of the prominent night sky lines at 5577
\AA\ and 6300 \AA.

The IRAF task {\tt rvsao} was used to obtain the redshift of the
galaxy by cross-correlation with several SDSS galaxy spectral
templates. Spectra from each grating setting were run through the
task separately. Each case returned a strong signal (R-value $>$6)
for an elliptical galaxy template with consistent redshift
values. Redshifts were verified by visual inspection of the spectra
using absorption features from Ca II H and K, the G-band, and Mg I
$\lambda$5172.7. No emission lines were present. Combining the
results from the two grating settings yielded a spectroscopic redshift
of $z=0.3242 \pm 0.0002$ for \lens.

ACT\,J2029 and \lens\ were also observed with the Near-Infrared Camera/Fabry–P\'erot Spectrometer (NICFPS) instrument on the 
ARC 3.5~m telescope at the Apache Point 
Observatory \citep{vincent03} with a $Ks$-band ($\lambda \sim 2.1$~$\mu$m) 
$5\sigma$ detection limit of $m=21.35$ mag. The data were reduced as described 
in Section 2.3 of \citet{menanteau13}. We use this photometry in our SED-fitting of the system, presented in Section \S\ref{sec:sedfit}.

\subsection{Pan-STARRS}
\begin{figure}
 \centering
 \includegraphics[width=0.99\columnwidth]{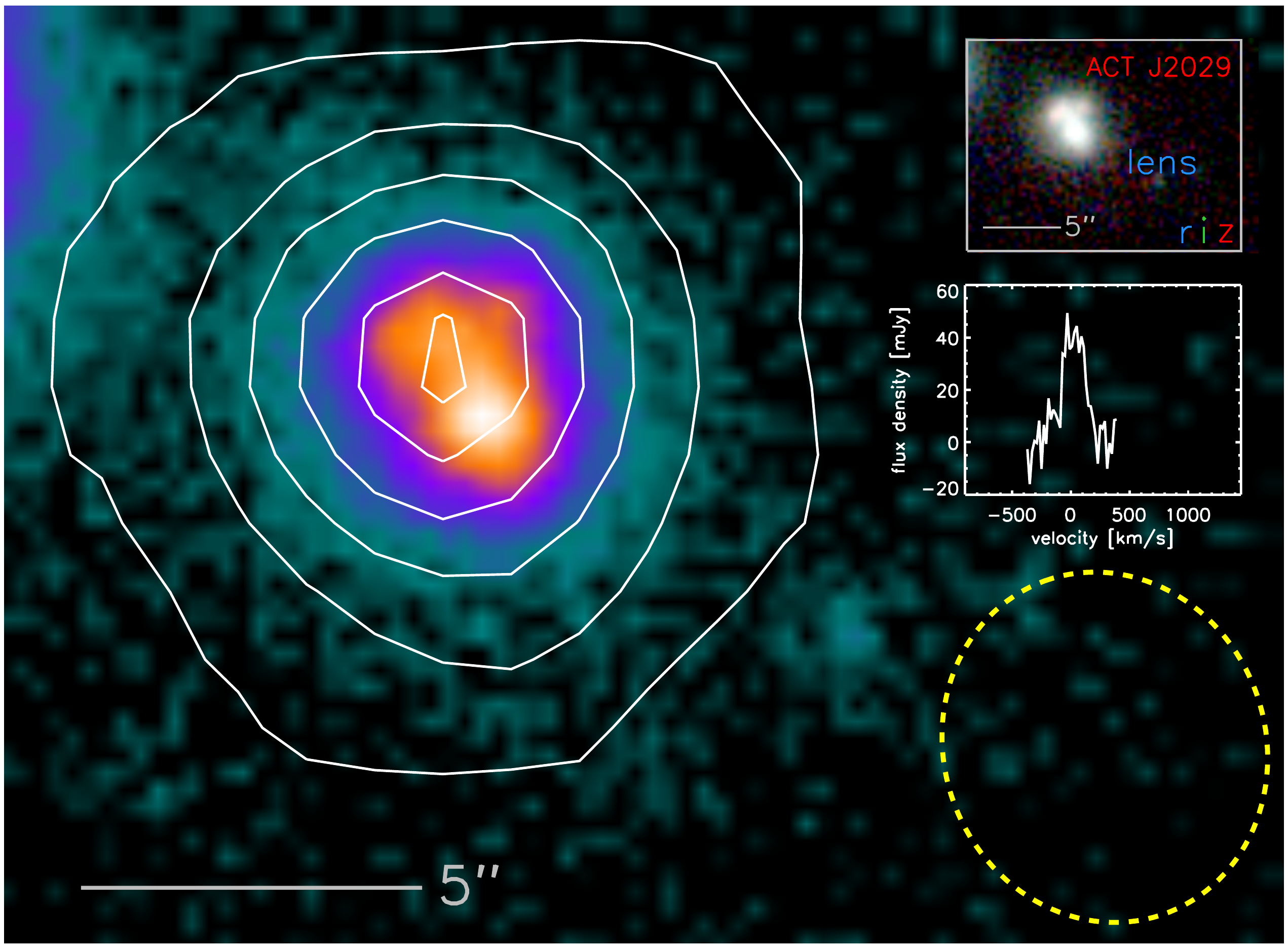}
 \caption{A Pan-STARRS $i$-band image of ACT\,J2029 and \lens\ with the CARMA contours overplotted (white). The ellipse at lower right represents the CARMA beamsize. An $riz$ color image and the positions of ACT\,J2029 and \lens\ is shown at upper right. The CARMA integrated emission line is shown in the middle right. The two sources in the $i$-band image have positions less than 0.5" apart and are spatially blended. However, what appears to be a lensed arc is clearly visible in the $riz$ image.\label{fig:lens}}
\end{figure}

In order to constrain the nature of the lensed system, we also make use of $grizy$ images from the Panoramic Survey Telescope and Rapid Response System (Pan-STARRS) Data Release 1 (DR1; \citealt{chambers16}), which became public on 2016 December 19. The $grizy$ filters reach 5$\sigma$ depths of 23.3, 23.2, 23.1, 22.3, 21.3 mag, respectively. 
An $i$-band image of the system is shown in Figure \ref{fig:lens}, with a color image located in the upper right inset. Our CARMA data are also overlaid on the $i$-band image. A S\'ersic profile fit to the image reveals slight excess emission to the northeast and southwest of the 
elliptical galaxy (see Appendix \ref{sec:appA}), suggestive of a background lensed galaxy. The color inset image in Figure \ref{fig:lens} appears to support this suggestion.


\subsection{IRAM 30m}
\label{sec:observations}
The IRAM 30m observations of ACT\,J2029 were carried out as part of the 2015 IRAM Summer School, over the nights of 2015 September 13 and 15 using the Eight MIxer Receiver \citep[EMIR, ][]{carter2012} mounted on the telescope. EMIR has four different bands, E090, E150, E230, and E330, spanning the 3, 2, 1, and 0.9 mm windows respectively. Each band provides \mbox{$\sim8$ GHz} of instantaneous, dual-polarization bandwidth.

The data were recorded using the Wideband Line Multiple Autocorrelator (WILMA), which provides a spectral resolution of \mbox{2 MHz}, corresponding to velocity resolutions of \mbox{$\sim6$}, 3, and \mbox{2 km\,s$^{-1}$} for the E090, E150, and E230 bands, respectively. Dual-frequency EMIR setups were used throughout the observations. We employed E090/E150 (\mbox{3/2 mm}) and E090/E230 (\mbox{3/1 mm}) configurations to search for CO transitions. In these configurations, each receiver has a bandwidth of \mbox{4 GHz}. The \mbox{3 mm} receiver was used as a reference and tuned to the previously detected \mbox{95 GHz} CO transition, while the \mbox{2 mm} receiver was tuned to \mbox{142.5 GHz} and \mbox{158 GHz}, and the \mbox{1 mm} receiver to \mbox{221 GHz} and \mbox{253 GHz}. Additional lines of high-density gas tracers CS(7--6), HCN(4--3), and HCO$^{+}$(4--3) were targeted with the Fast Fourier Transform Spectrometer (FTS). The FTS backend has a spectral resolution of \mbox{195 kHz}, corresponding to \mbox{$\sim 0.5$ km s$^{-1}$} for the E090 band. E090/E230 (\mbox{3/1 mm}) EMIR configurations were repeated with a slightly shifted 3 mm tuning at 97 GHz to better center on the dense gas tracers, and 1 mm frequency tunings at 97 GHz and 221/253 GHz, respectively.

The observations were carried out in wobbler switching mode, with a switching frequency of 1.5 Hz. Pointing was checked frequently and was found to be stable within 3". The calibration was done every 12 minutes using standard hot/cold-load absorber measurements. The data reduction was performed using the GILDAS software's \textsc{class} package. Scans with distorted baselines were excluded from the dataset ($\sim$22\%), while scans showing platforming (i.e., steep baseline jumps arising in the backend electronics) were corrected for this effect and included. Linear baselines were subtracted from the individual spectra, except for the \ci\ /CO(7--6) spectrum, for which we fit a first, third and fourth order polynomial (see Section \S\ref{subsec:dense_tracers} for reasons).

\section{Results}
\label{sec:results}

\subsection{The redshift of ACT\,J2029}
\label{subsec:redshift}

\begin{figure}
 \centering
 \includegraphics[width=0.99\columnwidth]{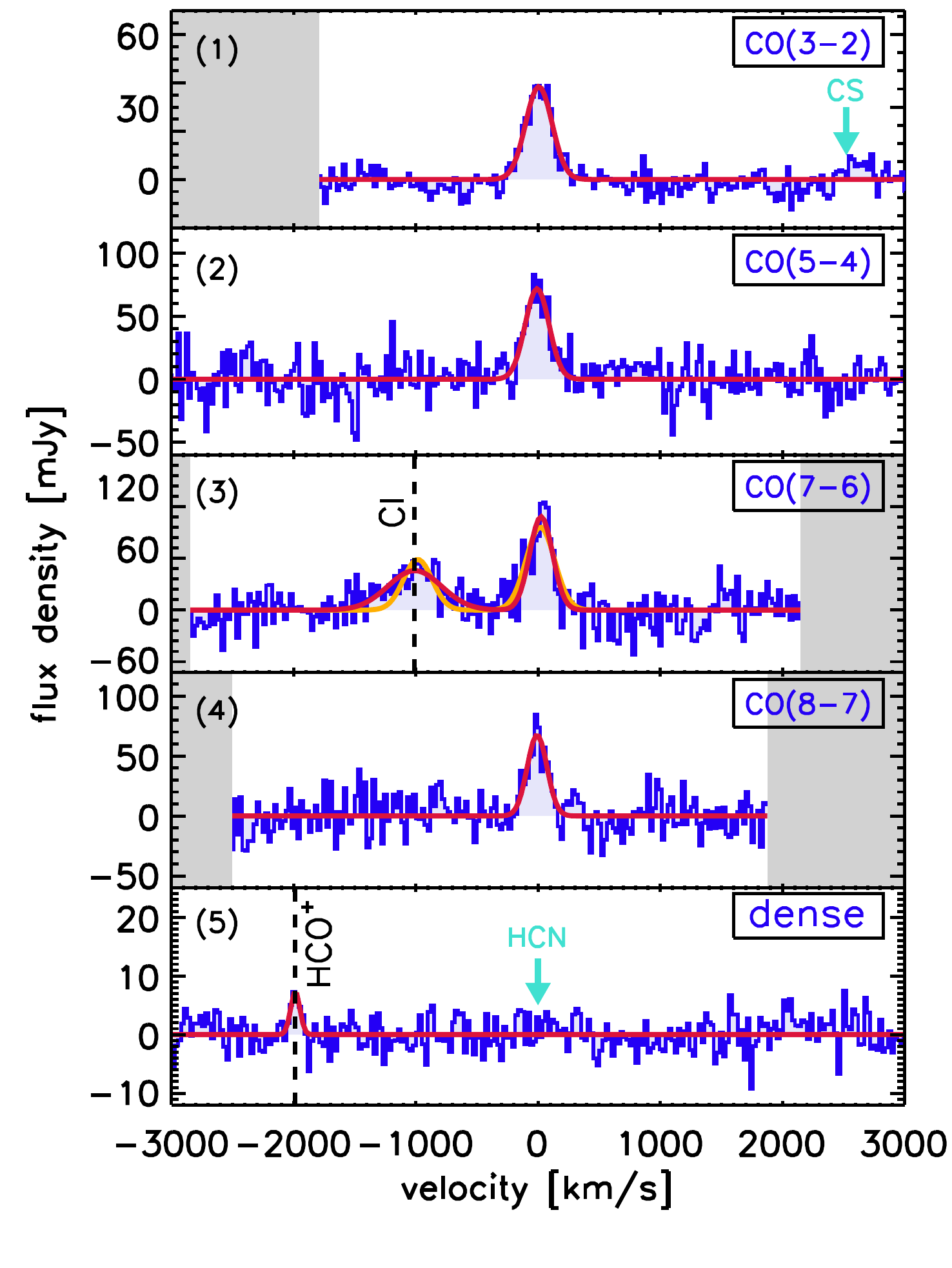}
 \caption{Observed $^{12}$CO and \ci\ transitions for ACT\,J2029, taken with the IRAM 30m. The 4 CO transitions observed with WILMA are $J$\,=\,3--2 (panel 1), 5--4 (panel 2), 7--6 (panel 3), and 8--7 (panel 4). Panels 1 and 5 display the upper limits on the other dense gas tracers and the tentative HCO$^{+}$(4--3) detection (taken with the FTS). Panel 3 also shows the \ci\ detection, with a linear baseline fit. The red line indicates the best fit Gaussian function from \textsc{class} and the gray shaded regions mark the limits of our data. The orange line in panel 3 represents a Gaussian fit to the \ci\ line fixed to width of CO(7--6). \label{fig:spec}}
\end{figure}

For our IRAM 30m observations we adopted an observing strategy similar to that developed by \citet{weiss2009} to determine the redshift of the object. The 3/2 mm setup was tuned to 95/142 GHz to confirm and improve the line shape of the 95 GHz feature and to test for redshifts \mbox{$z$\,=\,1.43} and \mbox{$z$\,=\,3.85}, resulting in a non-detection at $\sim$142 GHz. Subsequently, the 2 mm band was tuned to 158 GHz, where a second line was detected at 158.3 GHz. This uniquely identifies the 95.0 and 158.3 GHz line features as CO(3--2) and CO(5--4) at a redshift \mbox{$z$\,=\,2.64}. CO(7--6) and CO(8--7) lines at 221.6 and \mbox{253.2\,GHz} were also detected using the \mbox{3/1\,mm} (95/221\,GHz) and \mbox{3/1\,mm} (95/253\,GHz) setups, respectively. A redshift of \mbox{$z$\,=\,2.64002$\pm$0.00006} was derived by combining the centroids of all the CO lines using a variance-weighted average.
All four CO line profiles are very similar and well described by single Gaussian components, as shown in \mbox{Figure~\ref{fig:spec}}. Their line parameters are presented in \mbox{Table~\ref{tab:all_lines}}. The uncertainties associated with the integrated intensities are derived from the uncertainties in the free parameters of the Gaussian fits.


\subsection{Line properties}
\label{subsec:dense_tracers}
\ci($^{3}\mathrm{P}_{2}$$-$$^{3}\mathrm{P}_{1}$) was detected in conjunction with CO(7--6) at \mbox{222 GHz} using the EMIR \mbox{3/1 mm} (95/221 GHz) setup with the WILMA backend and applying a linear baseline correction. Like the CO line profiles, the \ci\ line profile is well described by a single Gaussian component, but with a FWHM that is more than double the CO line widths. 

In an attempt to constrain the robustness of the potentially larger FWHM, we employed a Monte Carlo approach to randomly select only 50\% and 75\% of the scans, subtract the baseline with a first order polynomial, and fit single Gaussians to the \ci\ and CO(7--6) lines. This process was repeated five times for each percentile. In all cases, the line width of the \ci\ profile remained extended relative to the CO(7--6). Additionally, we re-fit the profiles from the full data set using velocity bins ranging from 3 to 55\,km~s$^{-1}$. We find that the maximum variation from the determined \ci\ line width across all velocity bins is 67.4\,km\,s$^{-1}$ ($\sim$\,5\%), and from the measured peak flux density is 0.3 mJy (\textless1\%).
Finally, we tested subtraction of third and fourth order polynomial baselines before fitting a single Gaussian component to the \ci\ profile. This was repeated for four random sets of 75\% of the data, as well as for the full data set; all ten fits returned a \ci\ width larger than that of the CO lines. Therefore, the line width and flux determinations appear robust, suggesting real broadening of the \ci\ line compared to the CO lines.
Nevertheless, the larger \ci\ FWHM may be due to baseline effects and we caution that independent confirmation of the broadening is required for absolute certainty.
The implications of a wider line profile of \ci\ with respect to CO is discussed in Section \S\ref{subsec:enhanced_ci}. The profile of the \ci\ line is also very well fit by several two-component Gaussian fits, with a blueshifted component offset by $\sim$100-400 km $s^{-1}$ from a systemic component. The reduced $\chi^{2}$ of every two-component fit varies by less than $\Delta\chi^{2}\sim$0.01 relative to other two-component fits, and is always $\Delta\chi^{2}\sim$0.001 lower compared to the single-Gaussian fit. Therefore, we argue a two-component Gaussian fit is not sufficiently justified to replace a single Gaussian fit.

We also report a tentative detection of the HCO$^{+}$(4--3) line at 98 GHz in the 3 mm band using the FTS backend, with an integrated S/N ratio of 3.46. The line is well described by a single Gaussian profile, whose parameters are presented in \mbox{Table \ref{tab:all_lines}}. The line appears to be blueshifted with respect to the systemic (CO) redshift by more than 100 km s$^{-1}$, and also has a much narrower width than any of the CO transitions. Using Monte Carlo simulations that account for the peak flux of the HCO$^{+}$ line and the noise in our spectra, we find that in 10,000 random samples the probabilities of obtaining such a velocity offset and smaller line width with respect to our CO observations are less than 0.1\% and 0.01\%, respectively. Due to the low S/N ratio, however, it is difficult to draw robust conclusions, so we do not include this line in our analysis. From our observations, we also place strong upper limits on the emission of the HCN(4--3) and CS(7--6) lines, by requiring that line widths be similar to that of the CO emission lines and calculating $3\sigma$ integrated intensities. Limits on fluxes and luminosities are provided in \mbox{Table \ref{tab:all_lines}}.

\begin{deluxetable*}{lrcccccc}
\tablewidth{0pt}
\tablecolumns{6}
\tabletypesize{\footnotesize}
\tablecaption{All IRAM 30m line detections.\label{tab:all_lines}}
\tablehead{
\colhead{Line} & 
\colhead{$\nu_{obs}$ [GHz]} &
\colhead{FWHM [km/s]\tablenotemark{*}} &
\colhead{$S_{\nu}$ [mJy]\tablenotemark{*}} &
\colhead{I [Jy km/s]\tablenotemark{*}} &
\colhead{L$^\prime$\,[x10$^{10}$\,K~km~s$^{-1}$~pc$^{2}$]\tablenotemark{*}} &
\colhead{Integration time [s]} &
\colhead{EMIR Setup} 
}
\startdata
CO(3--2) &  94.997(2) & 270.1 $\pm$ 14.8 & 39.7 $\pm$ 2.8 & 11.3 $\pm$ 0.5 & 41.8 $\pm$ 1.9 & 9900 & E090 \\
CO(5--4) & 158.321(5) & 229.3 $\pm$ 25.2 & 71.7 $\pm$ 10.2 & 17.4 $\pm$ 1.5 & 23.1 $\pm$ 2.1 & 2820 & E150 \\
CO(7--6) & 221.59(2) & 231.3 $\pm$ 27.7 & 87.5 $\pm$ 13.1 & 21.4 $\pm$ 1.9 & 14.5 $\pm$ 1.3 & 7620 & E230 \\
CO(8--7) & 253.25(9) & 190.9 $\pm$ 22.7 & 72.8 $\pm$ 11.6 & 14.7 $\pm$ 1.5 & 7.6 $\pm$ 0.8 & 7140 & E230 \\
\ci\,(2--1) & 222.4(7) & 590.5 $\pm$ 98.8 & 37.6 $\pm$ 7.9\tablenotemark{b} & 23.5 $\pm$ 3.0 & 15.8 $\pm$ 2.0 & 7620 & E230 \\
CS(7--6) &  94.198\tablenotemark{a} & -- & $< 12.1$ & $< 1.0$ & $< 3.7$ & 24600 & E090 \\
HCN(4--3) &  97.392\tablenotemark{a} & -- & $< 8.9$ & $< 0.7$ & $< 2.5$ & 24600 & E090 \\
HCO$^{+}$(4--3) &  98.004\tablenotemark{a} & 94.2 $\pm$ 25.6 & 7.6 $\pm$ 2.8 & 0.7 $\pm$ 0.2 & 2.6 $\pm$ 0.7 & 24600 & E090 
\enddata
\tablenotetext{*}{All uncertainties are quoted at the 1$\sigma$ level and all values are uncorrected for lensing. Upper limits are robust to 3$\sigma$.}
\tablenotetext{a}{Expected redshifted frequency of the line, assuming $z$\,=\,2.64002 $\pm$ 0.00006.}
\tablenotetext{b}{The peak flux density derived with the line width fixed to that of the CO(7--6) line is $S_{\nu}$=51.5 $\pm$ 5.77 mJy.}
\end{deluxetable*}

As shown by \cite{solomon1992}, line luminosity can be expressed in terms of total line flux as:
\begin{equation} \label{eq:luminosity}
    L^{\prime}_\text{line} = 3.25 \times 10^{7}\ S_{line}\Delta v\ \nu^{-2}_{obs}\ D^{2}_{L} (1 + z)^{-3}  [\mbox{K~km~s$^{-1}$~pc$^{2}$}],
\end{equation}
where $S_{line}\Delta v$ is the velocity integrated flux in \mbox{Jy~km~s$^{-1}$}, $D_{L}$ is the luminosity distance in Mpc, and $\nu_{obs}$ is the observed central frequency of the line in GHz. The luminosities of all detected lines are derived using \mbox{Eq.~\ref{eq:luminosity}} and are listed in \mbox{Table \ref{tab:all_lines}}, along with all other line properties.

\section{Discussion}
\label{sec:disc}
The observations described above make it possible to characterize 
the molecular gas in ACT\,J2029 in an unusual level of detail ---
a useful opportunity given continuing uncertainties in exactly 
how DSFGs contribute to the cosmic histories of star formation 
and black hole accretion. We therefore proceed below to use 
the new data to assess the origin of ACT\,J2029's molecular 
excitation, determine whether it is a strongly lensed system, and 
interpret the anomalous ratios of its CO and \ci\ line widths and 
fluxes in light of our conclusions on those two points. 
\subsection{CO SLED indicative of an AGN}

\begin{figure*}
 \centering
 \includegraphics[width=0.99\textwidth]{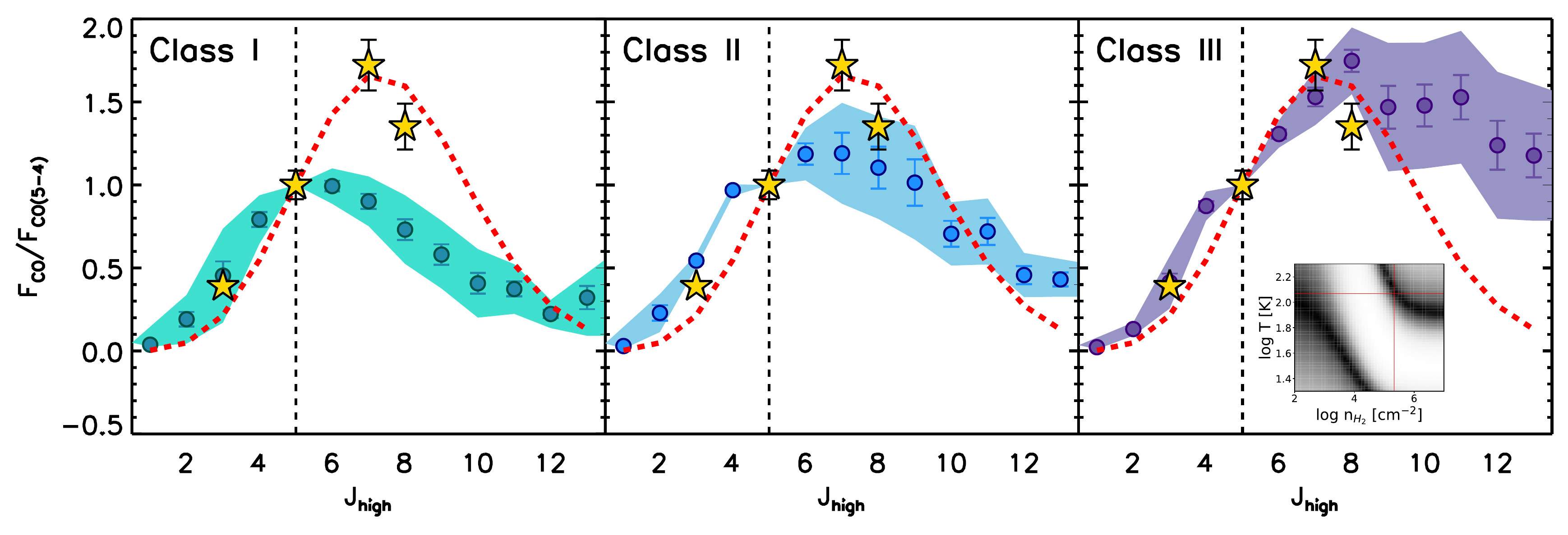}
 \caption[width=0.99\textwidth]{\label{fig:rosenberg}
The distribution of $^{12}$CO line fluxes, normalized to that of the $^{12}$CO(5--4) line, for the HerCULES sample of Class~I (star-forming objects, left panel), Class~II (starbursts and Seyferts, middle panel), and Class~III (ULIRGs and QSOs, right panel) objects \citep{rosenberg2015}. The filled circles represent the mean CO SLED, the colored shaded regions show their $\pm 1 \sigma$ deviations off the mean, and the error bars represent the uncertainty of the mean. The ACT\,J2029 fluxes are represented by the (yellow) stars.
The dashed red line in each panel represents the best RADEX fit to the observed CO fluxes, which corresponds to \mbox{$T_\text{kin}=117$ K}, \mbox{$n_{H} = 2 \times 10^5$ cm$^{-3}$} and \mbox{$N_{\text{CO}}/\Delta$v = $10^{16}$ cm$^{-2}$}(km s$^{-1}$)$^{-1}$. A T vs. n$_{H_{2}}$ probability plot is shown as an inset to the right panel, highlighting a range of other possible fits allowed by temperature-density degeneracies. The best RADEX fit is represented by the red crosshairs.}
\end{figure*}

Figure~\ref{fig:rosenberg} shows the $^{12}$CO spectral line energy distribution (SLED) constructed with our observations, which peaks at the $J$\,=7--6 transition. In the same plot, we compare our detections to the starbursts, ULIRGs, and AGN hosts that were observed as part of the \textit{Herschel} Comprehensive ULIRG Emission Survey (HerCULES; PI: van der Werf) by \citet{rosenberg2015}, normalized to the CO(5--4) line. It becomes immediately clear that ACT\,J2029 is most consistent with the Class~III objects from \citet{rosenberg2015}, which are mainly AGN hosts, although the observed CO transitions alone cannot definitively rule out a match to the Class~II objects. The CO consistency with the Class~II and Class~III objects suggests that the dense molecular gas in this source is exposed to the harsh radiation field of a central AGN \citep{vanderwerf2010,rosenberg2015,israel2015}.

In order to better characterize ACT\,J2029, we used the non-local thermal equilibrium code RADEX \citep{vandertak2007} to model the emitted CO intensities. A large parameter grid, $T_\text{kin} = 20 - 200$ K, $n_{\text{H}_2} = 10^2 - 10^7$ cm$^{-3}$, and $N_\text{CO}/\Delta$v = 10$^{17.9} - 10^{19.4}$ cm$^{-2}$ (km s$^{-1}$)$^{-1}$ (assuming $\Delta$v=10 kms$^{-1}$), was sampled, where $T_\text{kin}$ is the kinetic temperature, $N_{\text{CO}}$/$\Delta$v the column density per unit velocity gradient of CO, and $n_{\text{H}_2}$ the molecular gas number density. The best reproduction of the observed CO intensities is given by the parameters \mbox{$T_\text{kin} = 117$ K}, \mbox{$n_{\text{H}_2} = 2 \times 10^5$ cm$^{-3}$}, and $N_\text{CO}/\Delta$v = 3 $\times$ 10$^{18}$ cm$^{-2}$ (km s$^{-1}$)$^{-1}$.
This result is also consistent up to $J$=8--7 with the Class~II and Class~III objects from the \citet{rosenberg2015} sample, but is inconsistent with the Class~I objects: the fit agrees well with our data up to the $J$=8--7 transition, however it drops rapidly at higher-$J$. This is likely due to the use of single-component gas fits in our RADEX models, which are sufficient to trace the lower=$J$ transitions but insufficient to accurately trace the higher-$J$ transitions. More sophisticated modelling of the higher-$J$ transitions would likely require multiple-component gas fits and additional data.
The best-fit model is highlighted as a dashed red line in \mbox{Figure \ref{fig:rosenberg}}. It is important to note, however, that other similar fitting models exist due to degeneracies between \mbox{$T_\text{kin}$} and \mbox{$n_{\text{H}_2}$}. A probability density plot of the sampled models is shown as an inset plot in \mbox{Figure \ref{fig:rosenberg}}.

Several CO SLEDs with similar shapes have also been observed in shocked systems such as the central nuclear region of Centaurus A \citep{israel2014} and the shocked AGN-driven molecular outflow in NGC\,1266 \citep{alatalo2011,pellegrini2013,glenn2015}. In both cases, the CO SLED rises and peaks around CO(7--6), then decreases at higher-$J$. 
Such a pattern is also seen in both AGN hosts and shock hosts, however the transitions we have detected in ACT\,J2029 do not allow us to differentiate between these possibilities. In each of the first two shock-dominated sources noted above, an AGN is also present, mechanically or radiatively driving an outflow, which in turn produces the shocked line ratios in the CO SLED.

\citet{bothwell132} and \citet{spilker14} derive average 
DSFG CO SLEDs that look broadly similar to the SLED for 
ACT\,J2029; however, they have slightly shallower slopes 
(below their peak $J$) and less pronounced ``knees.'' In 
contrast, the CO SLED of Cloverleaf \citep{bothwell13}, a 
well-known AGN host, has a slope and knee that are in much
better agreement with those of ACT\,J2029.
The review by \citet{casey2014} on high-$z$ DSFGs characterizes the prevalence of AGN hosts among SMGs as high, but not universal, with 1/3 of sources showing signs solely of prolific star formation, and 2/3 requiring AGN. Therefore, an additional possibility exists to explain the shape of ACT\,J2029's CO SLED and, although no radio or X-ray detections of ACT\,J2029 exist, the CO SLED is suggestive of the source hosting an AGN. However, a definitive confirmation of the presence of an AGN will require higher resolution, deeper observations to characterize delensed morphology, or detections of X-rays or radio (jet/lobe) emission.

To assess whether ACT\,J2029 resembles the high-$z$ main sequence of star-forming galaxies, we compare its CO SLED to those of other star-forming objects in the literature. Using CO SLEDs from 4 BzK galaxies to obtain an average CO SLED from CO(1--0), CO(2--1), CO(3--2) and CO(5--4) transitions, \citet{daddi15} compare a BzK CO SLED to average CO SLEDs for local (U)LIRGs, SMGs and the inner disk of the Milky Way. Their results suggest ACT\,J2029 has a CO SLED very similar in shape to those of BzKs and SMGs up to the CO(5--4) transition. However, such a comparison extends only to CO(5–-4) whereas the “knee” of our data occurs at CO(7-–6), making it difficult to compare high-$J$ transitions where the nature of the excitation is likely to be determined.


\subsection{Lensing status and SED fitting}
\label{sec:sedfit}
As seen in Section \S\ref{sec:obs}, discrepant CO and optical redshifts are highly suggestive of gravitational lensing. With a secure DSFG redshift in hand, we are now in a position to quantify the lensing magnification. Although no lens model is available for our source, \citet{harris12} showed for a sample of star-forming galaxies that estimation of the lens magnification is possible using the line width and luminosity of the CO(1--0) transition (see also \citealt{bothwell13}, \citealt{goto15}; cf. \citealt{aravena16}, \citealt{sharon16}). Assuming a CO(3--2)/CO(1--0) luminosity ratio of 0.9$\pm$0.4 \citep{sharon16} and a CO(1--0)/CO(3--2) line width ratio of 1.15$\pm$0.06 \citep{ivison11} for ACT\,J2029, we estimate a lens magnification $\mu$=25$\pm$11 using Equation 1 of \citet{harris12}. Although a lens model is clearly needed to fully confirm any magnification and there is substantial uncertainty in our derived value, it becomes immediately clear that ACT\,J2029 is likely lensed.\footnote{
  We note that a CO(1--0) line width larger than that of CO(3--2) (e.g., as seen by \citealt{ivison11} for a small SMG sample) would lead to a lower magnification estimate.}

\begin{figure}
 \centering
 \includegraphics[width=0.99\columnwidth]{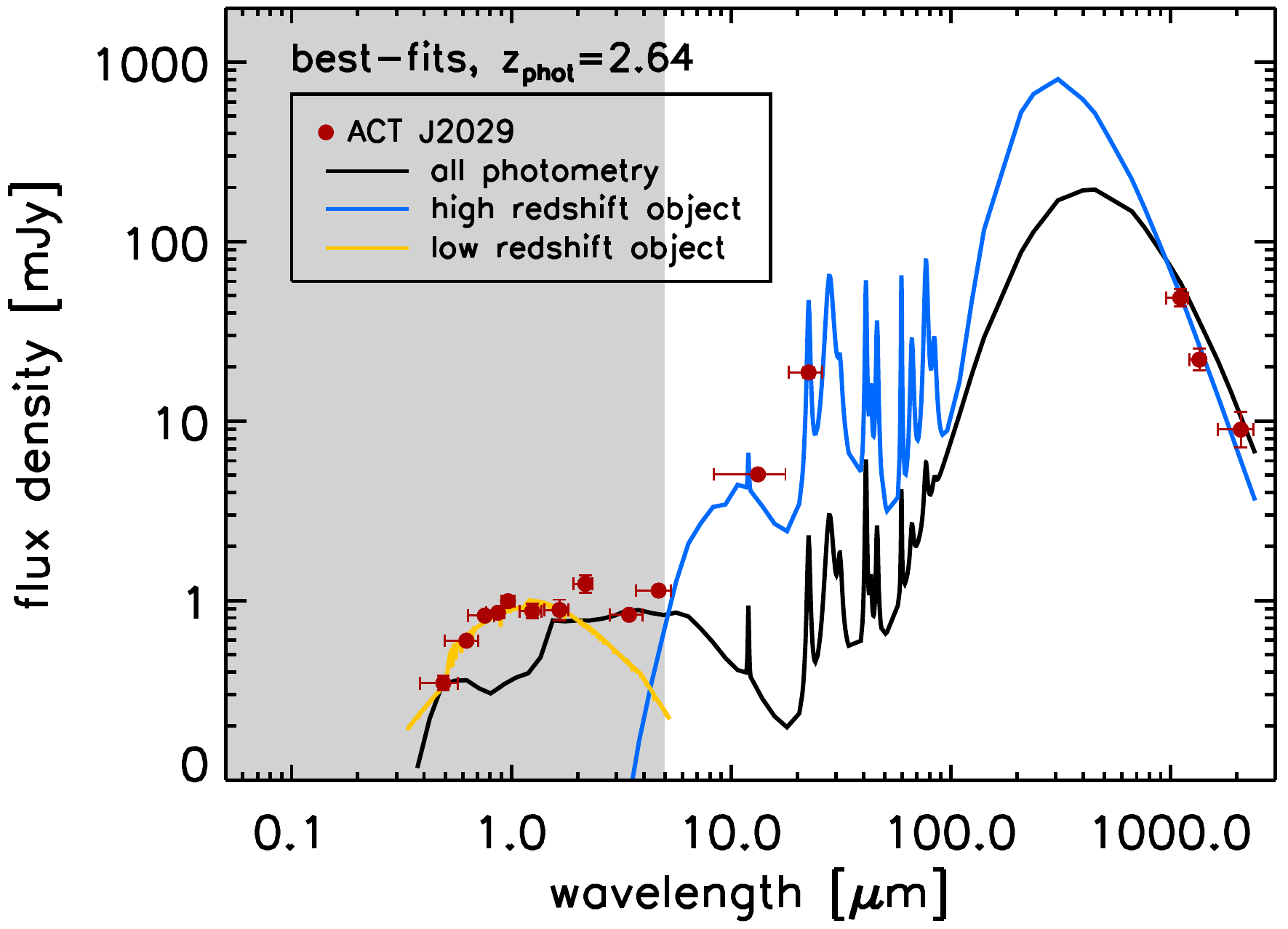}
 \caption{\label{fig:sed}
The best-fit SEDs for ACT\,J2029 (black and blue) and \lens\ (yellow), fixed to redshifts of $z$\,=\,2.64 and $z$\,=\,0.32, respectively. The gray shaded region represents wavelengths where the photometry is likely contaminated by both the high-$z$ object and the foreground object. The black line illustrates the best fit ACT\,J2029 SED making use of all available photometry, while the blue line uses the same data but with upper limits to all photometry blueward of the \textit{WISE} W3 band (ie., the shaded region). The yellow line shows the best-fit SED to \lens, applying upper limits to all the photometry in the shaded region. A \citet{calzetti2000} extinction law was preferred for all fits.}
\end{figure}

We also make use of Pan-STARRS \citep{chambers16} $grizy$ filters, 2MASS \citep{skrutskie06} J- and H-band, NICFPS Ks-band, all \textit{WISE} \citep{wright10} bands, and our ACT photometry to fit an SED to ACT\,J2029, using the \textsc{Hyperz} code developed by \citet{bolzonella}. The optical photometry was obtained by using the SExtractor \citep{bertin96} software in dual-image mode on each of the Pan-STARRS images, whilst the NIR photometry was taken from the 2MASS and \textit{WISE} All-Sky Data Releases, respectively. 
We run \textsc{Hyperz} with the standard library templates, which include nebular emission lines \citep{coleman80,kinney96,fioc97,silva98,chary01,bc03,polletta07,michalowski10}, a fixed photometric redshift of $z$\,=\,2.64 to match our derived spectroscopic redshift, and a dust column with A$_{\text{v}}$\,$\in$\,[0.0-3.0] mag following either a \citet{calzetti2000} starburst or a \citet{prevot84} Small Magellanic Cloud extinction law. Each template has been fit (by itself) to our data twice, once with the \citet{calzetti2000} law and once with the \citet{prevot84} law. We first use all of our SED constraints and conclude that none of the templates we used can correctly reproduce the photometry of ACT\,J2029. However, as discussed in the previous section, the most likely hypothesis for our object is that it is lensed by a foreground galaxy at $z$\,=\,0.3242, and therefore the photometry at shorter wavelengths, namely shortward of the \textit{WISE} W3 band, is contaminated by the foreground galaxy. If we use only \textit{WISE} W3, W4 and ACT photometric constraints while setting upper limits to all photometry blueward of $\approx$1$\mu$m, we find a better fit with reduced $\chi^2 \sim$ 1.63. We also fit a low-redshift SED using only upper limits on all photometry blueward of $\approx$1$\mu$m and a fixed redshift of $z$\,=\,0.32. The derived SEDs are presented in Figure~\ref{fig:sed}. 
It becomes obvious that the optical data are well fit by the low-redshift object, while the \textit{WISE} W3 and W4 bands are elevated and likely contaminated by both objects.
We conclude that the elevated \textit{WISE} points suggest that the NIR fluxes are indeed contaminated by the foreground source, \lens, making it impossible to derive a robust SED fit from which one can estimate global properties for ACT\,J2029. Accurate modeling of the contaminating fluxes would likely require higher-resolution photometry to fully characterize the low-redshift foreground galaxy.


\subsection{The enhanced \ci\ flux and velocity}
\label{subsec:enhanced_ci}

\begin{figure}
 \centering
 \includegraphics[width=0.99\columnwidth]{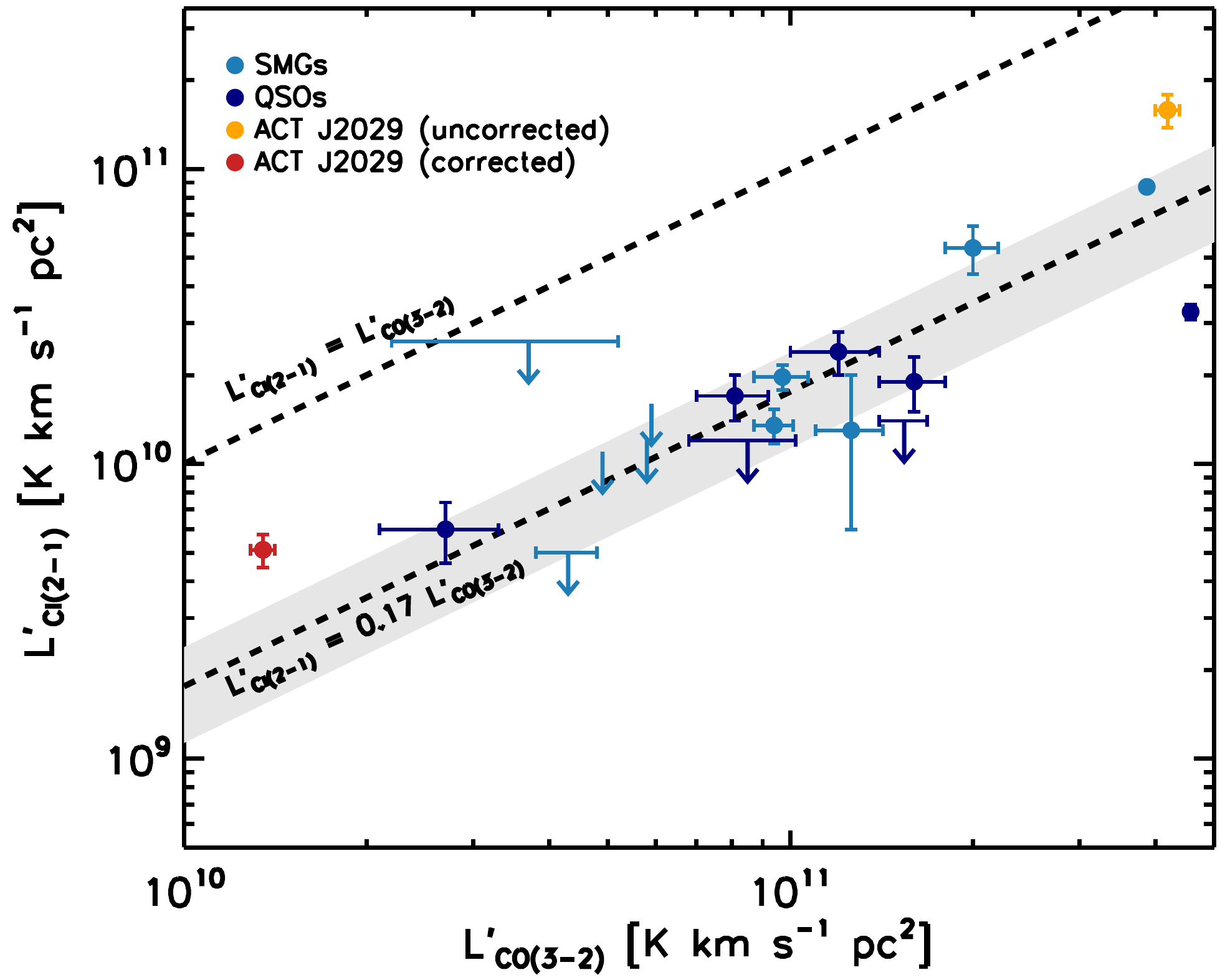}
 \caption{\label{fig:lcolci} The $L'_{\text{CI}(2-1)}/L'_{\text{CO}(3-2)}$ ratio of ACT\,J2029 (yellow and red, uncorrected and corrected for lensing, respectively) is overplotted with other lensed and unlensed high-redshift QSOs (navy blue) and SMGs (teal) from \citet{walter2011}. All \citet{walter2011} $L'$ values are uncorrected for lensing. The average $L'_{\text{CI}(2-1)}/L'_{\text{CO}(3-2)}$ line is overplotted, with the 1$\sigma$ standard deviation limits shaded in gray. Even taking magnification into account, ACT\,J2029 has a $L'_{\text{CI}(2-1)}/L'_{\text{CO}(3-2)}$ ratio substantially higher than other high-redshift galaxies. Classifications of SMGs and QSOs were taken from the \textsc{D\'IGAME}\footnote{http://www.digame.online/} catalogs.}
\end{figure}

The mean value of the \ci\,(2--1)/CO(3--2) luminosity ratios in \citet{walter2011} shows that in most SMGs the \ci\,(2--1) line has a luminosity $(0.17 \pm 0.06)L'_{\text{CO}(3-2)}$, consistent for both galaxies that host quasars and those that do not. ACT\,J2029 shows an excess of \ci\,(2--1) emission compared to CO(3--2), with \mbox{$L'_{\text{CI}(2-1)}/L'_{\text{CO}(3-2)} = 0.378 \pm 0.051$}, twice the mean value for other high-redshift sources. This excess is significant at greater than a 2$\sigma$ level and is shown in Figure~\ref{fig:lcolci}, which we compare to the lensed and unlensed objects presented in \citet{walter2011}. 

Most local and high-$z$ sources do not show substantial \ci\ flux enhancements \citep{walter2011,alaghbad-zadeh2013,israel2015}, but there are some marked exceptions. \citet{pellegrini2013} show that the local AGN-driven molecular outflow host NGC\,1266 displays enhanced \ci\ emission compared to $^{12}$CO, as does the nuclear region of Centaurus A \citep{israel2014}. In both cases, it is argued that the enhanced \ci\ emission is due to the shocked gas chemistry in the nuclear regions where the line emission originates (in both cases, this scenario is supported by the CO SLED). ACT\,J2029 could be a similar Class~III object hosting an AGN in its center, with shocked dense gas contributing to an enhanced \ci/CO line ratio.

The difference in the \ci\ and CO linewidths of ACT\,J2029 is also unusual. In all objects where resolved velocity information is available for both \ci\ and CO, the linewidths of the two lines are consistent (\citealt{walter2011}, and references therein). In Section \S\ref{subsec:dense_tracers}, we showed that the \ci\,(2--1) line is more than twice as wide as the CO lines, suggesting that the source of the \ci\ and CO emission cannot be identical.

If we combine the large linewidth difference with the enhanced \ci\,, an intriguing possibility suggests itself, namely, that we are observing differential lensing of a compact nuclear region of this source, which hosts an AGN-driven molecular outflow. Lensing preferentially enhances compact regions \citep{hezaveh2012}, so it is possible that the foreground lens (\lens) has magnified the nuclear region of ACT\,J2029, including outflowing gas traced by \ci\ and CO. \citet{rawle2014} used the Submillimeter Array and the Karl G. Jansky Very Large Array to map the molecular lines of HLS0918, a lensed submillimeter galaxy at $z$\,=\,5.2430. In this case, they were able to spatially differentiate multiple velocity components in the molecular gas, including a very broad (VB) component consistent with an outflowing region. \citet{rawle2014} show that the ratio of \ci\,(2--1)/CO(7--6) in the VB component is larger than those seen in all other regions, suggestive of enhanced \ci\,(although the velocity structures in HLS0918 are complex, such that degeneracies arise when fitting the components in the blended \ci\ and CO(7--6) lines). It is possible that ACT\,J2029 exhibits a similar CI-enhanced outflow region, while lacking the complex velocity structures that are observed in the \citet{rawle2014} source.


\section{Conclusions}
This work presents multi-transition $^{12}$CO and \ci\,(2--1) measurements of ACT\,J2029, a lensed sub-millimeter galaxy found by the ACT survey. Following up on earlier line detections by the LMT and CARMA, we confirm a source redshift of \mbox{$z$\,=\,2.640}. To summarize our work:

\begin{itemize}
\item The CO(3--2), CO(5--4), CO(7--6), CO(8--7), and \ci\,(2--1) transitions are detected at high significance, but only upper limits on the emission of the high-density gas tracers CS(7--6) and HCN(4--3) are obtained. Additionally, we also report a tentative detection of the HCO$^{+}$(4--3) line.

\item We confirm a redshift of $z$\,=\,2.64 for ACT\,J2029, based on IRAM 30m observations of multiple CO lines.

\item We construct a CO SLED starting from the four detected CO transitions. Based on a comparison with the star-forming, star-bursting, and AGN-hosts from the HerCULES survey, the CO SLED of ACT\,J2029 is consistent with a Class~II or Class~III object, i.e., with a starburst, ULIRG, or powerful AGN host.

\item Non-LTE modeling suggests that ACT\,J2029 is more consistent with a Class~III AGN host object, characterized by a \mbox{$T_\text{kin} = 117$ K}, \mbox{$n_{\text{H}_2} = 2 \times 10^5$ cm$^{-3}$}, and \mbox{$N_\text{CO}/\Delta$v = 3 $\times$ 10$^{18}$ cm$^{-2}$} (km s$^{-1}$)$^{-1}$ modulo non-trivial parameter degeneracies.

\item We provide convincing evidence to support the lensing status of ACT\,J2029 by (i) spectroscopically confirming the redshift of a foreground source inconsistent with that of ACT\,J2029, and whose position lies less than 0.5" away, and (ii) estimating a magnification factor of $\mu$$\approx$25 for ACT\,J2029 via the fiducial relation of \citet{harris12}.

\item The velocity width of the \ci\ line appears to be 
substantially larger than what is seen in all CO transitions, 
and the $L'_{\text{CI}(2-1)}/L'_{\text{CO}(3-2)}$ ratio appears
to be larger than what is typically seen in lensed and un-lensed 
SMGs and QSO hosts. The latter discrepancy would be in agreement 
with what has been observed in shocked systems, such as 
Centaurus\,A and NGC\,1266. If confirmed, the large \ci\ width 
and enhanced \ci\ emission could be explained by differential 
lensing, in which a shocked, centrally concentrated outflow 
(traced by the enhanced \ci) has been preferentially magnified 
compared to the larger scale molecular gas (traced by CO).


\end{itemize}

\acknowledgements
The authors thank Richard Tunnard, Thomas Greve, Am\'elie Saintonge, Jackie Hodge, Yashar Hezaveh, and Paul van der Werf for excellent conversations helping to interpret the \ci\ emission, and Tony Wong for help with CARMA data reduction.
We also thank the organizers of the IRAM 2015 Summer School, from which the IRAM 30m observations were acquired, and an anonymous referee for comments that helped improve the paper.
KA is supported through Hubble Fellowship grant \hbox{\#HST-HF2-51352.001} awarded by the Space Telescope Science Institute, which is operated by the Association of Universities for Research in Astronomy, Inc., for NASA, under contract NAS5-26555.
This work was supported by the U.S. National Science Foundation through awards AST-0955810 to AJB and AST-0408698 and 
AST-0965625 for the ACT project, along with awards 
PHY-0855887 and PHY-1214379. Funding was also provided by Princeton 
University, the University of Pennsylvania, and a Canada Foundation for 
Innovation (CFI) award to the University of British Columbia. ACT operates 
in the Parque Astron\'omico Atacama in northern Chile under the auspices of 
the Comisi\'on Nacional de Investigaci\'on Cient\'ifica y Tecnol\'ogica de Chile 
(CONICYT). Computations were performed on the GPC supercomputer at the 
SciNet HPC Consortium. SciNet is funded by the CFI under the auspices of 
Compute Canada, the Government of Ontario, the Ontario Research 
Fund—Research Excellence, and the University of Toronto.
Some of the observations reported in this paper were obtained with the
Southern African Large Telescope (SALT). We would like to thank Encari
Romero Colmenero for her care in carrying out the SALT observations.

\newpage
\appendix

\section{NIR sersic fit and potential lens morphology}
\label{sec:appA}

\begin{figure*}[h]
\epsscale{0.85}
\plotone{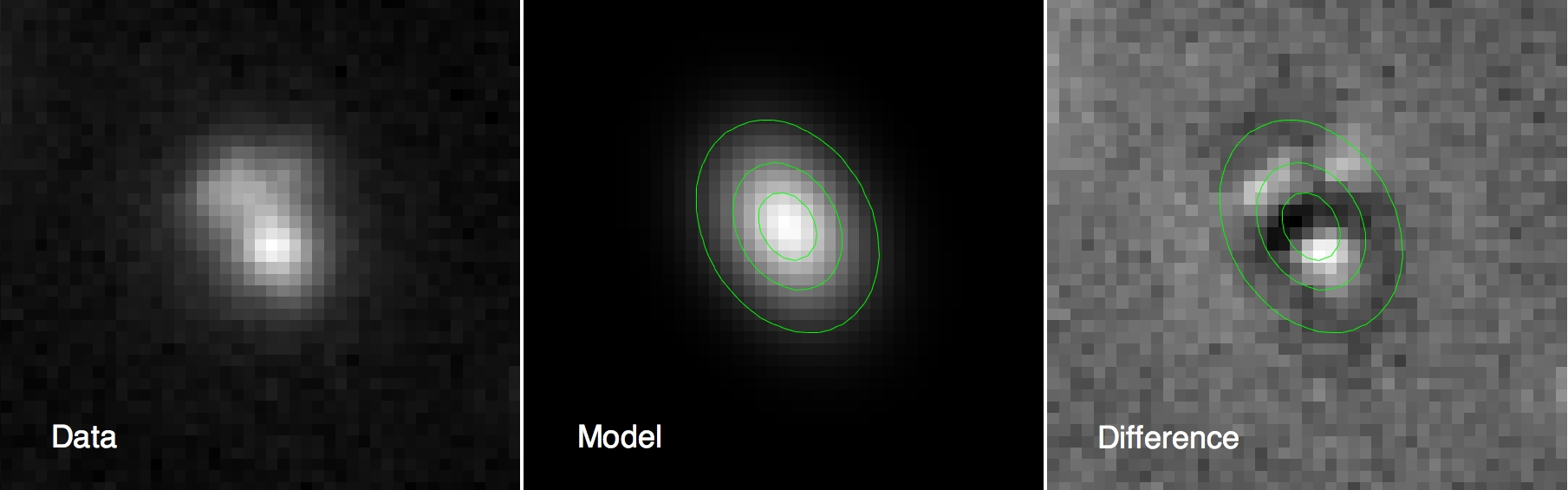}
\caption{The {\sc Galfit} analysis of the foreground galaxy in the Pan-STARRS $i$-band image. The left panel shows the $i$-band image, centered on ACT\,J2029, the middle panel shows the {\sc Galfit} S{\'e}rsic profile fit of the foreground galaxy, and the right panel shows the difference between the left and middle panels (ie., the data minus the the model). NIR excess can be seen towards the northeast and southwest of the center.
\label{fig:sersic}}
\end{figure*}

We used the two-dimensional profile fitting tool {\sc Galfit} 
\citep{peng+02,peng+10} to study the morphology of the lensing 
galaxy in the Pan-STARRS $i$-band image. A single component, elliptical S{\'e}rsic profile provided a 
reasonable fit to the morphology resulting in an effective radius of 
$1.54^{\prime\prime} \pm 0.01^{\prime\prime}$, a S{\'e}rsic index of 
$0.59\pm0.01$, and an axial ratio of $0.75 \pm 0.01$ at a position 
angle of $29^\circ \pm 1^\circ$. The small value of the 
S{\'e}rsic index indicates a fairly compact galaxy. The interesting 
feature of the fit is the pattern of residuals that appear in the 
difference image (data minus model) shown in Figure~\ref{fig:sersic}. There is an 
arc-like excess toward the north/northeast about $1.5^{\prime\prime}$ 
from the fitted center of the galaxy and another, more compact, excess 
$\sim$0.4$^{\prime\prime}$ southwest of the center. We estimate the 
summed intensity of these features to be at least one magnitude 
fainter than the lensing galaxy. These results are suggestive (but not 
definitive) of NIR emission from the background lensed galaxy. Higher 
resolution imaging will be needed to confirm this. 

\end{document}